\documentclass[epj]{svjour}
\usepackage{latexsym,amssymb,amsfonts,mathrsfs,color,graphicx,amsmath}% etc
\usepackage[usenames,dvipsnames]{xcolor}

\DeclareMathOperator{\atanh}{atanh}
\DeclareMathOperator{\asec}{asec}

\begin{document}

\newcommand{\vhh}[1]{\hat{\vec{#1}}}
\newcommand{\eh}{\hat{\vec{e}}}
\newcommand{\hsy}[1]{\hat{\vec{#1}}}
\newcommand{\ten}[1]{\vec{\sf #1}}
\newcommand{\RCh}{R_{\text{Ch}}}
\newcommand{\RSq}{R_{\text{Sq}}}
\newcommand{\NMD}{N_{\text{MD}}}
\newcommand{\vdd}[1]{\frac{d}{dt}\vec{#1}}
\newcommand{\vdh}[1]{\frac{d}{dt}\hat{\vec{#1}}}

\newcommand{\hphi}{\hat{\vec{\varphi}}}
\newcommand{\hrho}{\hat{\vec{\rho}}}
\newcommand{\hz}{\hat{\vec{z}}}
\newcommand{\hx}{\hat{\vec{x}}}
\newcommand{\hy}{\hat{\vec{y}}}
\newcommand{\ephi}{e_\varphi}
\newcommand{\erho}{e_\rho}
\newcommand{\ez}{e_z}
\newcommand{\htheta}{\hat{\vec{\theta}}}
\newcommand{\vf}{\bar{v}_f}
\newcommand{\HDD}{H_\text{2D}}
\newcommand{\HDDD}{H_\text{3D}}
\newcommand{\const}{\text{const}.}

\newcommand{\refEq}[1]{Eq.~(\ref{Eq:#1})}
\newcommand{\refEqu}[1]{Eqs.~(\ref{Eq:#1})}
\newcommand{\refFig}[1]{Fig.~\ref{Fig:#1}}
\newcommand{\refFigu}[1]{Figs.~\ref{Fig:#1}}
\newcommand{\refSec}[1]{Sec.~\ref{Sec:#1}}
\newcommand{\vecin}[2]{#1 \!\cdot\! #2}

\newcommand{\vout}{\vec{v}_\text{out}}
\newcommand{\vin}{\vec{v}_\text{in}}
\newcommand{\viout}{\vec{v}_\text{i,out}}
\newcommand{\viin}{\vec{v}_\text{i,in}}

\newcommand{\anm}[1]{\textsl{\textcolor{Bittersweet}{[#1]}}}

\renewcommand{\phi}{\varphi}

\title{Periodic and Quasiperiodic Motion of an Elongated Microswimmer in Poiseuille Flow}
%\subtitle{Do you have a subtitle?\\ If so, write it here}

\author{Andreas Z\"{o}ttl \and Holger Stark% etc
% \thanks is optional - remove next line if not needed
%\thanks{\emph{Present address:} }%
}                     % Do not remove
%
%\offprints{}          % Insert a name or remove this line
%
\institute{Institut f\"{u}r Theoretische Physik, Technische Universit\"{a}t Berlin, Hardenbergstrasse 36, 10623 Berlin, Germany \\ \email{andreas.zoettl@tu-berlin.de}}

\date{Received: \today / Revised version: date}
% The correct dates will be entered by Springer
%
\abstract{
We study the dynamics of a prolate spheroidal microswimmer in Poiseuille flow for different flow geometries. When moving between two parallel plates or in a cylindrical microchannel,
the swimmer performs either periodic swinging or periodic tumbling motion.
Although the trajectories of spherical and elongated swimmers are qualitatively similar, the swinging and tumbling frequency strongly depends on the aspect ratio of the swimmer.
In channels with reduced symmetry the swimmers perform quasiperiodic motion which we demonstrate explicitely for swimming in a channel with elliptical cross section.
\PACS{
      {47.63.Gd}{Swimming microorganisms}   \and
      {47.15.G-}{Low-Reynolds-number (creeping) flows}   \and
      {02.30.Ik}{Integrable systems}
     } % end of PACS codes
} %end of abstract
\maketitle

\section{Introduction}
\label{intro}
Swimming on the micron scale has attracted a lot of attention among physicists,
since they try to understand the fascinating strategies, microorganisms employ to
overcome the constraints of low Reynolds number hydrodynamics \cite{Purcell77,Lauga09}.
On the other hand, artificial microscopic swimmers or active particles with various locomotion mechanisms have been constructed recently also with the goal of using controllable
environments to study principal properties of their diffusive and collective motion, as well as locomotion under external fields such as gravity or shear flow \cite{Paxton04,Dreyfus05,Gauger06,Howse07,Jiang10,Palacci10,Thutupalli11,Volpe11,Theurkauff12}.
The experiments are accompanied by detailed theoretical studies
analyzing hydrodynamic  (see e.g.\ \cite{Simha02,Hatwalne04,Hernandez-Ortiz05,Llopis06,Saintillan07,Underhill08,Baskaran09,Nash10,Rafai10}),
stochastic (see e.g.\ \cite{Peruani06,Tailleur08,Ginelli10,Enculescu11}), chemotactic (see e.g. \cite{Keller71,Taktikos12}),
optical \cite{Pototsky12}
or thermophoretic \cite{Golestanian12}
effects on the dynamics of active particles.

In nature, microorganisms move in an aqueous environment and often have to respond to 
fluid flow like microscopic plankton in the sea \cite{Guasto12}, pathogens in the blood stream \cite{Uppaluri12}, or sperm cells in the fallopian tubes when swimming towards the egg \cite{Riffell07}.
Rheotaxis of sperm cells \cite{Bretherton61} or bacteria \cite{Marcos12}
 allow directed motion along flow gradients.
 Bottom-heavy microorganisms show stable orientations in flow resulting in layering
of phytoplankton in the ocean  \cite{Durham09} or hydrodynamic focussing
in Poiseuille flow \cite{Kessler85,Pedley87}. Neutrally buoyant microswimmers, however, 
follow periodic trajectories when swimming in Poiseuille flow \cite{Zilman08,Zoettl12,Uppaluri12}.
Finally, microswimmers in time-dependent flow easily show chaotic motion
and non-trivial transport phenomena \cite{Torney07,Khurana11}.

In our previous Letter we have studied the  dynamics of spherical microswimmers in cylindrical Poiseuille flow and identified two basics swimming states, an upstream oriented swinging
motion around the centerline and tumbling similar to passive particles \cite{Zoettl12}. In this article, we extend our work and analyze the motion of elongated microswimmers, as they occur more
frequently in nature, subject to Poiseuille flow in microchannels with different geometries. In our study, we will assume that gravity or other external forces are not present or not relevant.
Our swimmers do not exhibit active rotation (as in Ref.~\cite{Wittkowski12}), so their local reorientation in flow is the same
as for passive elongated particles. The hydrodynamic reorientation in low Reynolds number flow 
depends on both the local flow vorticity and  the local strain rate in the fluid. They both determine 
the periodic tumbling motion of passive ellipsoidal particles in flow commonly known as Jefferey 
orbit \cite{Jeffery22}. The strain rate does not influence the orientation of spherical particles and 
was therefore not considered in our Letter \cite{Zoettl12}. 

Now, in contrast to passive particles, microswimmers also change their position in flow due to their active motion. While rigid passive particles just follow streamlines during rotation \cite{Bretherton62},
active particles cross streamlines because of their self-propulsion. In this article we 
demonstrate that microswimmers with elongated shape still perform the characteristic
swinging and tumbling motion in steady Poiseuille flow as observed for spherical
swimmers. The motion is periodic in planar Poiseuille flow or in cylindrical tubes
with spherical cross section. Interestingly, the motion becomes quasiperiodic even for
spherical swimmers when the flow cross section becomes elliptical.

In the following we first introduce the equations of motion in section\ \ref{sec.equ}, 
discuss the swimmer trajectories for different flow cross sections in section\ \ref{sec.Poiseuille},
and conclude with final remarks in section \ref{sec.concl}.

\section{Equations of motion} \label{sec.equ}

\begin{figure}[htb]
\begin{center}
\includegraphics[width=0.75\columnwidth]{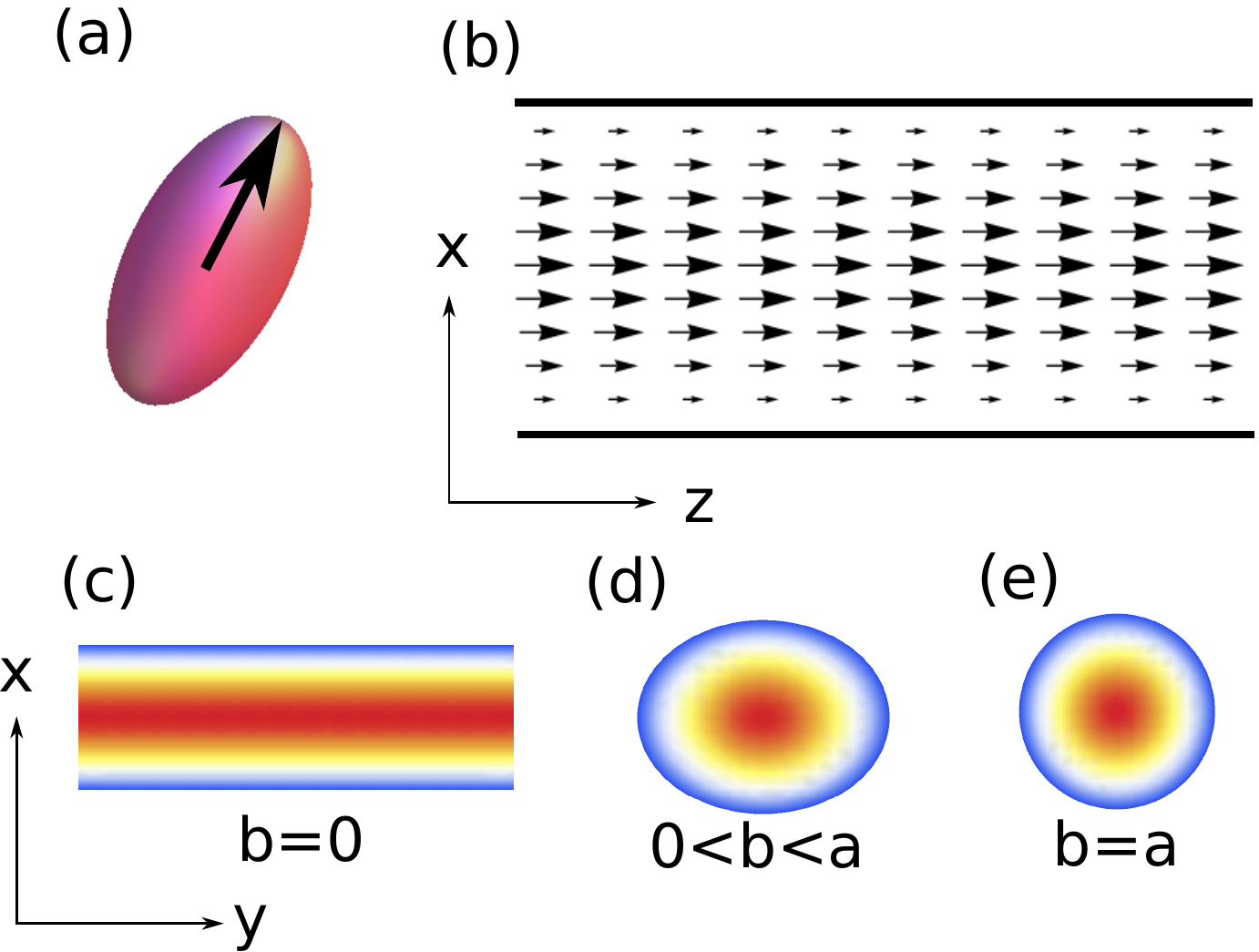}
\end{center}
\caption{
(a) Spheroidal microswimmer with aspect ratio $\gamma$
and orientation $\vec{e}$ (black arrow).
(b) Sketch of the Poiseuille flow profile 
in the $x$-$z$ plane.
(c)-(e): Sketch of the channel cross sections ($x$-$y$-plane)
with semimajor axis $a^{-1}$ and  semiminor axis $b^{-1}$ 
and color-coded flow strength for (c)
two parallel and infinitely extended plates ($b=0$)
or a channel with (d) elliptical ($0<b<a$) and (e) circular ($b=a$) cross section.
}
\label{Fig:geometry}
\end{figure}

We study the dynamics of an elongated  or prolate spheroidal microswimmer with aspect ratio $\gamma$
in a steady laminar flow $\vec{v}_f(\vec{r})$ at low Reynolds number.
In bulk, the swimmer moves with a constant velocity $\vec{v}_{0} = v_0\vec{e}$,
where $\vec{e}$ is the 
swimming direction along the major axis of the swimmer [\refFig{geometry}(a)].
Low Reynolds number also means that we can neglect any inertia of the microswimmer.
In addition, we consider the case of large rotational Peclet number,
where rotation due to flow vorticity exceeds rotational diffusion.
Furthermore, the microswimmer should be  small compared to the lateral extension
of the imposed flow in the microchannel
and should not disturb the Poiseuille flow field.
Finally, we assume that the swimmer stays away from bounding walls so that we can neglect steric
and hydrodynamic interactions between swimmer and walls which otherwise play
an important role \cite{Hill07,Berke08,Llopis10,Drescher11,Zoettl12}.
Under these conditions, a passive particle simply follows the streamlines with its center-of-mass
velocity at position $\vec{r}$ given by $\vec{v}(\vec{r}) = \vec{v}_f(\vec{r})$.
Its orientation moves on a Jefferey orbit, where the total angular velocity
for a prolate spheroidal particle reads  \cite{Jeffery22}
\begin{equation}
\vec{\Omega}(\vec{r}) = \frac{1}{2}\vec{\Omega}_f(\vec{r}) + G \vec{e} \times [\tens{E}(\vec{r})\cdot\vec{e}].
\label{Eq:Om2}
\end{equation}
Here we have introduced the local flow vorticity
\begin{equation}
\vec{\Omega}_f(\vec{r}) = \nabla \times \vec{v}_{f}(\vec{r})
\label{Eq:Om1}
\end{equation}
and the local strain rate
\begin{equation}
\tens{E}(\vec{r}) = \frac{1}{2}\left[ \nabla\vec{v}_f(\vec{r}) +(\nabla\vec{v}_f(\vec{r}))^T \right].
\label{Eq:E}
\end{equation}
The geometry factor  $G=\frac{\gamma^2-1}{\gamma^2+1}$ with $G\in [0,1)$
depends on the aspect ratio $\gamma$.
For spherical swimmers, where $\gamma=1$ and $G=0$,
only  the flow vorticity $\vec{\Omega}_f$ contributes to the swimmer's angular velocity.
An \textsl{active} particle with intrinsic swimming speed $v_0$
and moving along its orientation $\vec{e}$ assumes the total velocity
\begin{equation}
\vec{v}(\vec{r}) = \vec{v}_f(\vec{r}) +v_0\vec{e},
\label{Eq:v2}
\end{equation}
so that it can easily cross streamlines.
Since we do not assume an intrinsic rotation rate of the swimmer,
 the angular velocity is the same as for a passive particle [\refEq{Om2}].
Using \refEqu{Om2} and (\ref{Eq:v2}) the dynamics for position $\vec{r}(t)$ and orientation $\vec{e}(t)$ of the swimmer follows
\begin{equation}
\vdd{r} = \vec{v}(\vec{r}), \quad
\vdd{e} = \vec{\Omega}(\vec{r})   \times \vec{e}.
\label{Eq:EOM1}
\end{equation}

We consider a Poiseuille flow
along the $\hz$-direction,
\begin{equation}
\vec{v}_f(x,y) = v_{f} [1-(ax)^2-(by)^2]\hz,
\label{Eq:PF1}
\end{equation}
where we use Cartesian coordinates ($x,y,z$) with the coordinate
basis $(\hx,\hy,\hz)$ and
$v_{f}$ as the maximum flow velocity in the center of the channel. % and $a$,$b=$ const.
For $b=0$, \refEq{PF1} describes the flow between two parallel and infinitely extended plates.
For $b=a$ the flow occurs in a cylindrical channel,
and for $0<b<a$ the  channel has a general elliptical cross section,
where $a^{-1}$ is the semimajor axis and $b^{-1}$ the semiminor axis [\refFig{geometry}(b-d)].
The flow vorticity is given by
\begin{equation}
\vec{\Omega}_f(x,y) = 2v_{f} ( a^2x \hy - b^2y \hx )
\end{equation}
and
the strain rate is 	
\begin{equation}
\tens{E} = -v_{f} \left( a^2x\hx\hz + b^2y\hy\hz +a^2x\hz\hx +b^2y\hz\hy \right).
\end{equation}

In the following we use rescaled units,
$ax \rightarrow x$, $by \rightarrow \bar{b}y$, $az \rightarrow z$,
and $t/t_0 \rightarrow t$ with $t_0 = (av_0)^{-1}$.
We also introduce the dimensionless flow speed $\vf = v_f / v_0$
and the aspect ratio of the channel cross section, $\bar{b} = b/a$.
The third parameter of the system is the aspect ratio $\gamma$ of the swimmer.

We note that the general configuration space for the swimmer dynamics is
the product
$\mathbb{R}^3\otimes\mathbb{S}^2$ of position space ($\vec{r}\in \mathbb{R}^3$)
and unit sphere ($\vec{e}\in\mathbb{S}^2$).
Since we only consider unidirectional flow, i.e.\ flow in z direction,
the dynamics of the swimmer does not depend on the $z$ coordinate,
and the relevant phase space becomes four-dimensional, $\mathbb{R}^2\otimes\mathbb{S}^2$.
The dimension of the relevant phase space is further reduced when  symmetric flow geometries are used,
such as planar or cylindrical flow.

\section{Swimmer dynamics in Poiseuille flow} \label{sec.Poiseuille}

In the following we study the deterministic dynamics of a prolate spheroidal microswimmer
in Poiseuille flow for several geometries.
We start with the motion between two parallel plates ($\bar{b}=0$),
continue with a cylindrical microchannel 
with  circular cross section ($\bar{b}=1$), and
finally discuss the motion in a channel 
with general elliptical cross section ($\bar{b}<1$).

\subsection{Planar Poiseuille flow}
\label{Sec:planar}
First, we discuss the motion of a microswimmer in planar Poiseuille flow,
$\vec{v}_f = (1-x^2)\hz$,
with flow vorticity
$\vec{\Omega}_f = 2\vf x$ and strain rate $\tens{E} = - \vf x(\hx\hz+\hz\hx)$.
For the position of the swimmer we employ Cartesian coordinates,
$\vec{r} = x\hx+y\hy+z\hz$, and
we expand the orientation vector $\vec{e}$ in the Cartesian basis using angular coordiantes
$\Psi$ and $\Theta$ such that
$\vec{e} = e_x(\Psi,\Theta)\hx + e_y(\Theta)\hy + e_z(\Psi,\Theta)\hz$ with
\begin{equation}
e_x = -\cos\Theta\sin\Psi, e_y = \sin\Theta, e_z=-\cos\Theta\cos\Psi.
\label{Eq:eh}
\end{equation}
Here $\Psi$  indicates the orientation relative to the $z$-axis
and $\Theta \neq 0$ the orientation out of the $x$-$z$-plane.
We note that $\Psi=0$, $\Theta=0$ corresponds to perfect upstream orientation
and  $\Psi=\pi$, $\Theta=0$ to downstream orientation, respectively.
Due to translational  symmetry of the flow field in $y$-direction the relevant
configuration space
of the swimmer reduces to three dimensions,
$ \{ x, \Psi, \Theta  \} \in \mathbb{R}\otimes\mathbb{S}^2 $.

Using \refEqu{EOM1} and~(\ref{Eq:eh}), the coupled equations of motion for $(x,\Psi,\Theta)$ read
\begin{equation}
\begin{split}
\dot{x} &=  -\cos\Theta\sin\Psi,\\
\dot{\Psi} &= \vf x ( 1-G\cos 2\Psi ), \\
%\dot{\Psi} &= \vf x\left[ 1-G\cos 2\Psi   \right] \\
\dot{\Theta} &=  \vf x\, \frac G 2 \sin 2 \Theta \sin 2\Psi.
\label{Eq:Plane3DG}
\end{split}
\end{equation}
Whereas the first equation describes cross-streamline motion of the microswimmer,
the second and third
equations 
are the same for passive ellipsoidal particles but now the position $x$ is time-dependent.
When the solutions $x(t)$, $\Psi(t)$ and $\Theta(t)$ of
\refEqu{Plane3DG} are known, $y(t)$  and $z(t)$ follow by integration,
\begin{equation}
\begin{split}
y(t) &= y(0) + \int_0^t\!\! \sin\Theta(t')\mathtt{d}t', \\
z(t) &= z(0) + \int_0^t\!\! \left[ \vf\left( 1-x(t')^2 \right) -\cos\Theta(t')\cos\Psi(t') \right]\mathtt{d}t'.
\label{Eq:yz}
\end{split}
\end{equation}

\begin{figure}[htb]
\includegraphics[width=0.9\columnwidth]{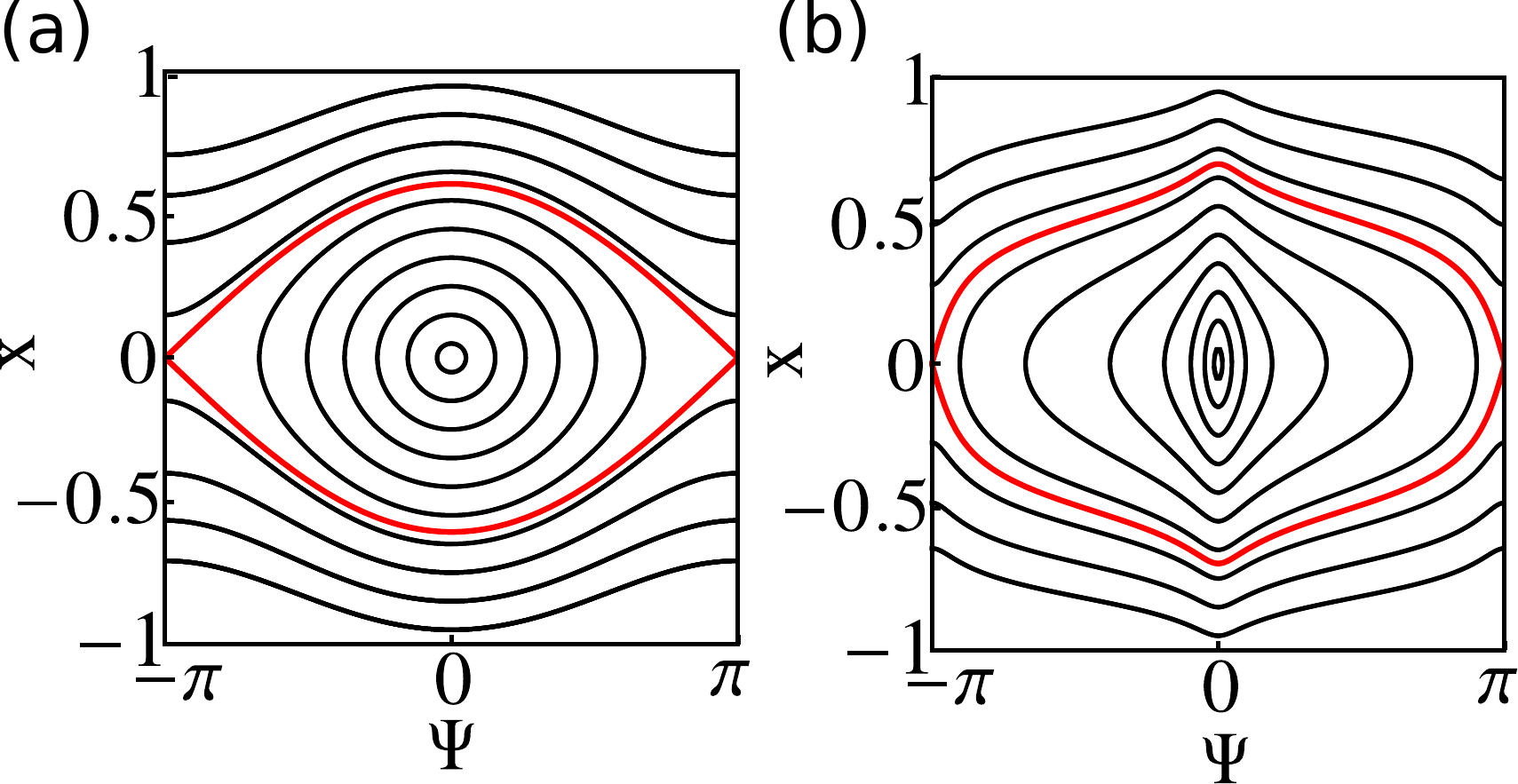}
\caption{$\Psi$-$x$ phase portraits for $\vf=10$ 
for a spherical swimmer with $\gamma = 1$ or $G=0$ (a) and an elongated 
swimmer with  $\gamma = 5$ or $G=0.923$ (b).
The red line is the separatrix deviding swinging and tumbling motion.
}
\label{Fig:xPsi2D}
\end{figure}

\subsubsection{Two-dimensional motion}
Before discussing the full three-dimensional dynamics of the swimmer,
we consider first the two-dimensional case. When
$\Theta=0$, the swimmer only moves in the $x$-$z$-plane and
the relevant phase space becomes two-dimensional,
 $ \{x,\Psi\} \in \mathbb{R}\otimes\mathbb{S} $.
 The dynamics of position $x$ and orientation $\Psi$
 follows from
\begin{equation}
\begin{split}
\dot{\Psi} &= \vf x ( 1-G\cos 2\Psi  ), \\
\dot{x} &=  -\sin\Psi.
\label{Eq:1}
\end{split}
\end{equation}
For a spherical swimmer $(G=0)$
we recover the case treated in Ref.~\cite{Zoettl12}.
The dynamic equations are
 $\dot{\Psi} =\vf x$ and $\dot{x} =  -\sin\Psi$,
which gives the pendulum equation, $\ddot{\Psi} + \omega_0^2\sin\Psi = 0$,
where $x$ plays the role of angular velocity
 and  $\omega_0^2=\vf$.
The corresponding constant of motion is the Hamiltonian  \cite{Zoettl12}
\begin{equation}
H = \frac 1 2 \vf x^2 + 1 -\cos\Psi.
\label{Eq:H2DG0}
\end{equation}

For elongated swimmers, whose dynamics is governed by \refEqu{1}, we are
also able to identify a constant of motion,
\begin{equation}
C_0 = \frac 1 2 \vf x^2 + 1 -g_0(\Psi; G)
\label{Eq:H2D}
\end{equation}
with 
\begin{equation} 
g_0(\Psi; G) =  \frac{\atanh (\sqrt{2G/(1+G)}\cos\Psi)}{\sqrt{2G(1+G)}}
\end{equation}
where $\atanh$ is the inverse function of the hyberbolic tangent.
We derive $C_0$ in Appendix \ref{Sec:COM2D}.
Note that $C_0 \rightarrow H$ for $G \rightarrow 0$.
For each swimmer trajectory the constant of motion is
uniquely defined 
by the initial conditions $\Psi(0)$ and $x(0)$.
We note that steric interactions of the swimmer with the channel wall
modify $C_0$.
After contact with the wall at $|x|=1$ the swimmer leaves the wall with upstream orientation 
$\Psi=0$ and the constant of motion becomes $C_{0} = \frac{\bar{v}_f}{2} + 1 - g_{0}(0;G)$  $ \approx 
\frac{\bar{v}_f}{2} + \frac{G}{3}(1-\frac{7}{5}G) $ for small $G$.

Figure~\ref{Fig:xPsi2D} shows the  $x$-$\Psi$ phase portrait
for $\vf=10$.
For $G=0$ (or $\gamma=1$) it corresponds to the mathematical pendulum (a).
The phase portrait for elongated swimmers looks qualitatively the same (b): Closed
and open periodic trajectories exist devided by a separatrix.
The trajectories are the curves
\begin{equation}
x^2 (\Psi) = \frac{2}{\bar{v}_{f}} [ g_0(\Psi;G) -1 + C_0  ].
\label{Eq:xPsiTraj}
\end{equation}
The closed trajectories correspond to swinging motion of the microswimmer in
Poiseuille flow as discussed in Ref.\ \cite{Zoettl12}.
Since $\Psi=0$ means  upstream motion, the swimmer swings
around the centerline of the channel while swimming upstream.
Because of the upstream orientation,
the flow vorticity always reorients the swimmer
towards the centerline.
After crossing the center at $x=0$ with maximum angle $\Psi_{\text{max}}$,
the sign of the vorticity changes and the swimmer is reoriented  back
towards the centerline.
So, the swimmer experiences a periodic swinging motion
around the centerline, 
in full analogy to the oscillations of a pendulum.
In the channel frame, the swimmer either moves upstream ($\dot{z}<0$) for suffieciently small
flow strengths $\bar{v}_{f}$ whereas it drifts downstream ($\dot{z}>0$) when $\bar{v}_{f}$
becomes large.
Since
\begin{equation}
\dot{z}(t) = v_z(t) + e_z(t) = \vf[1-x(t)^2] -\cos\Psi(t),
\label{Eq:zdot}
\end{equation}
the swimmer moves upstream for $v_z < -e_z$ 
or $\vf \lesssim 1$.
Perfect upstream swimming in the center of the channel
corresponds to the stable fixed point  $(\Psi=0,x=0)$.

 \begin{figure}[htb]
\begin{center}
\includegraphics[width=\columnwidth]{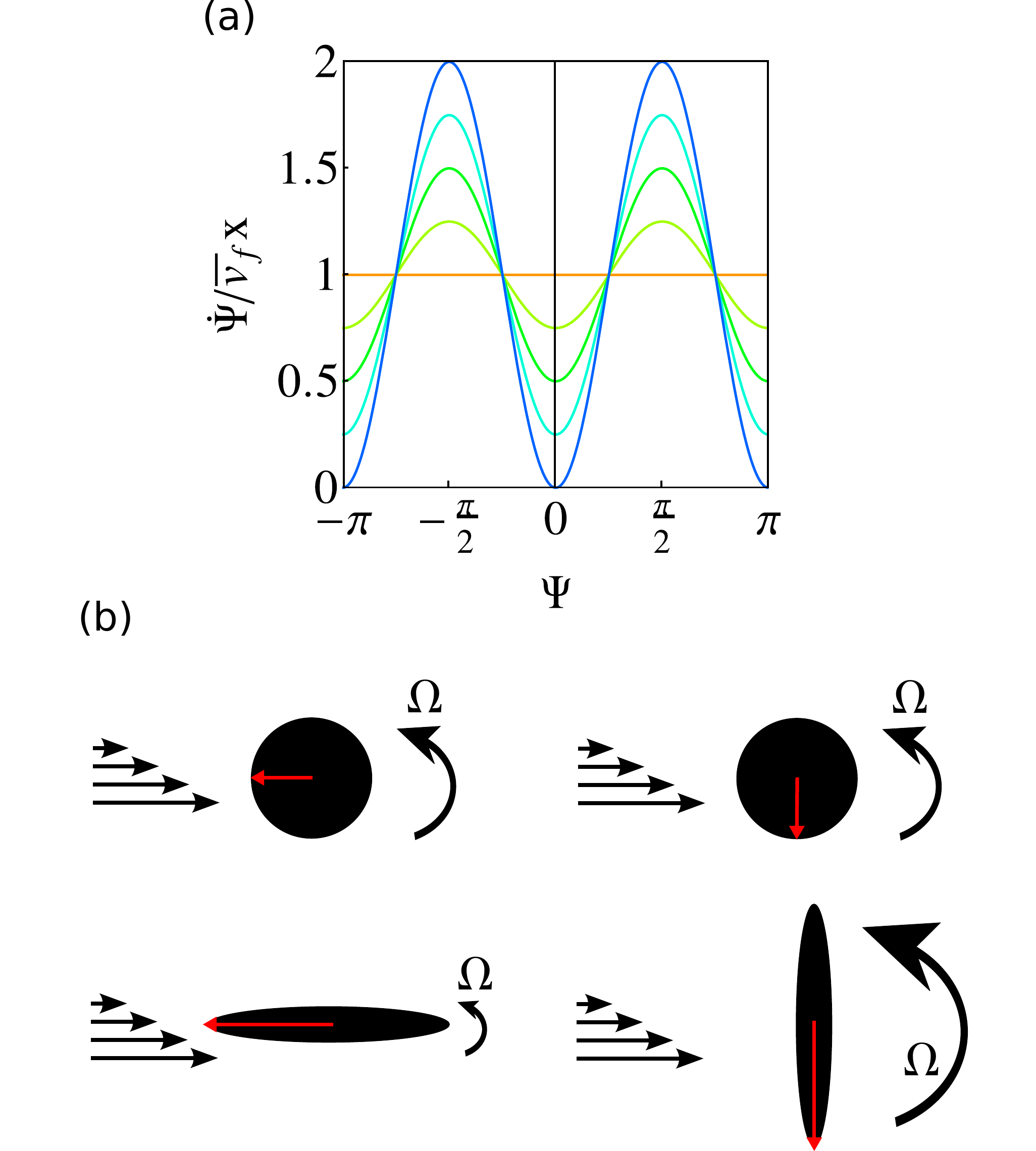}
\end{center}
\caption{
(a)  The angular velocity $\Omega_y = \dot{\Psi}$ at position $x$ and given flow strength $\bar{v}_{f}$
plotted versus swimmer orientation $\Psi$
for several shape parameters $G$;
orange line: $G=0$; light green: $G=0.25$; dark green: $G=0.5$;
light blue: $G=0.75$; dark blue: $G=1$.
(b): Sketch of the angular velocities of spherical
and elongated swimmers oriented parallel 
and perpendicular to the flow.
}
\label{Fig:rotation}
\end{figure}

\begin{figure}[htb]
\includegraphics[width=.99\columnwidth]{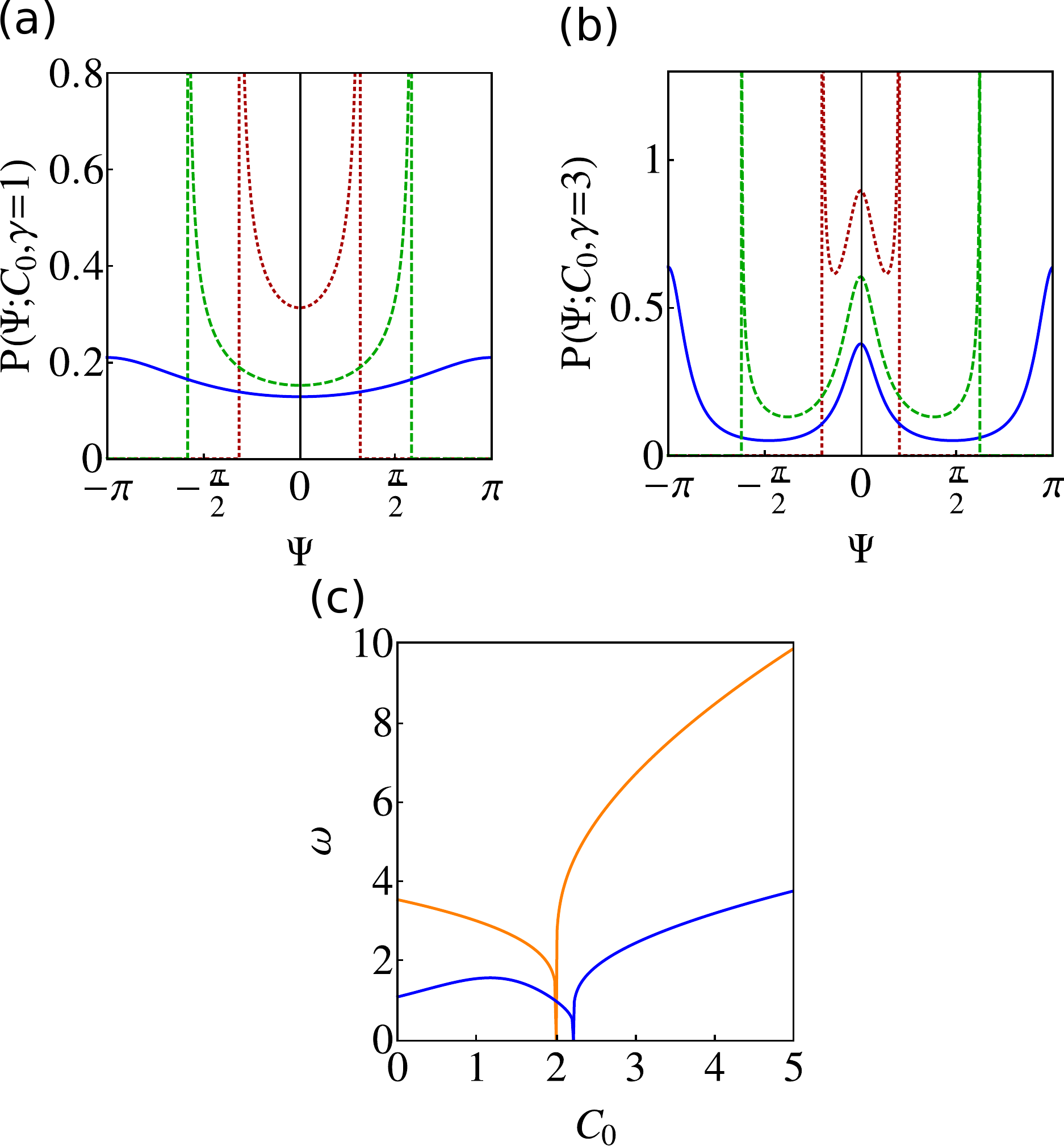}
\caption{
(a)-(b) Probability distributions $P(\Psi)$ for single trajectories with 
initial conditions $\Psi(0)=0$ and $\rho(0) = 0.3$ (red), $\rho(0) = 0.5$ (green)
and  $\rho(0) = 0.8$ (blue)
for spherical (a) and elongated (b) particles with $\gamma=3$ 
($G=0.8$).
(c) Angular frequencies $\omega$ for swinging and tumbling
motion for $\vf=10$ for a spherical swimmer (orange)
and an elongated swimmer with $\gamma=5$ ($G=0.923$) (blue) 
plotted versus the constant of motion $C_0$.
The frequency for small oscillations is $\omega \approx \sqrt{\bar{v}_f(1-G)}$.
}
\label{Fig:rotation2}
\end{figure}

\begin{figure*}[htb]
\includegraphics[width=\textwidth]{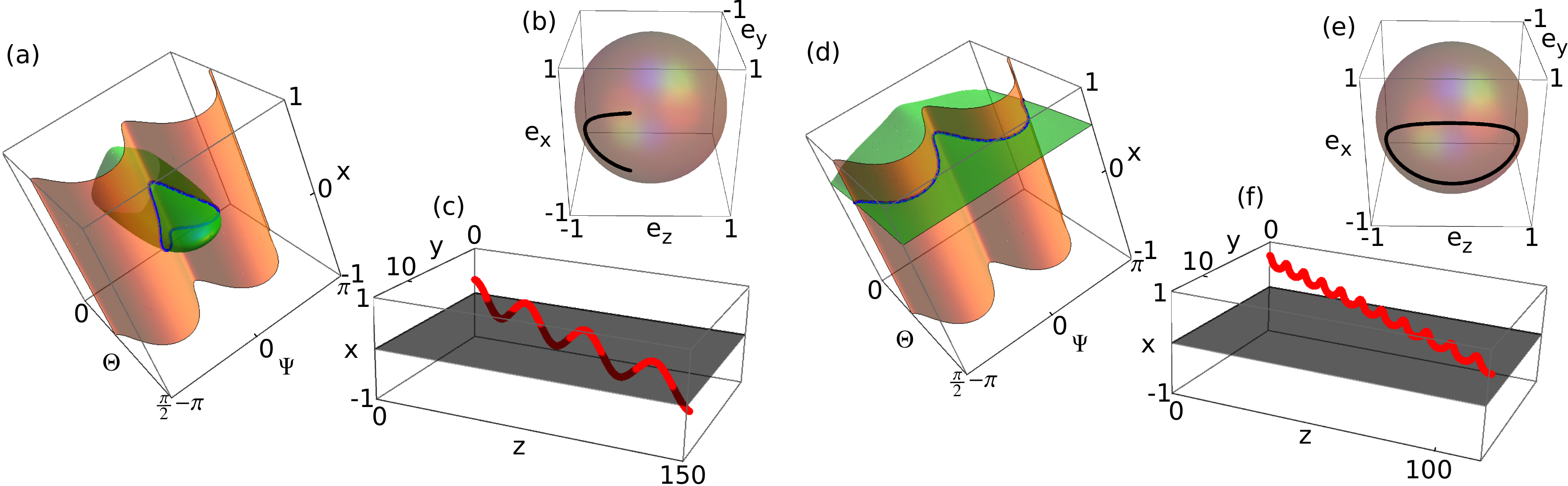}
\caption{  Swinging (a-c) and tumbling (d-e) motion of a spheroidal swimmer in flow between two parallel plates.
The flow strength is $\bar{v}_{f} = 10$ and the aspect ratio  of the swimmer is $\gamma=3$ 
(meaning $G=0.8$).
The initial conditions are $\Psi(0)=0$ and $\Theta(0)=\pi/5$
for both cases and  $x(0)=0.3$ for swinging and $x(0)=0.7$ for tumbling motion.
(a,d): Relevant three-dimensional phase space ($x,\Psi,\Theta$).
The intersection of the two constants of motion $\kappa_1$ (orange)
and $\kappa_2$ (green)  gives the periodic 
trajectory (blue intersection curve).
While $\kappa_1=2.64$ is the same for swinging (a) and tumbling (d),
$\kappa_2=1.14$ for swinging and $\kappa_2=2.86$ for tumbling motion.
(b,e): Periodic orbits of the orientation vector $\vec{e}(t)$ which moves on the unit sphere. 
For swinging motion it is a half ellipse (b) and for tumbling motion a closed curve (e).
(c,f): Swimmer trajectory $\vec{r}(t)$ in real space for swinging (c) and tumbling (f).
The midplane is shown in grey and the trajectories are red.
}
\label{Fig:3DGraphs}
\end{figure*}

When the initial state of the swimmer lies outside the separatrix, it tumbles in the flow.
The vorticity is too strong for the swimmer to reach the centerline. 
This is in analogy to the circling motion of a pendulum.

Swinging and tumbling motion are devided by the separatrix
(red curves in \refFig{xPsi2D}).
The trajectories $x(\Psi)$ [\refEq{xPsiTraj}] 
correspond to tumbling
when a position $x$ with  $\Psi=\pi$ exists or when $g_0(\pi;G) -1 + C_0 >0$.
Otherwise they are swinging trajectories.
The limiting case is the separatrix which for a 
fixed $\vf$ and $G$ is given by
\begin{equation}
x^\ast(\Psi^\ast;G,\vf) = \pm \sqrt{\frac{2}{\vf}\left[ g_0(\Psi^\ast;G)-g_0(\pi;G)  \right]}.
\end{equation}
The separatrix possesses the constant of motion
%The constant of motion of the separatrix is given by 
$C_{0}^{\ast} = 1-g_0(\pi;G) \approx 2 - \frac{G}{3}(1-\frac{7}{5}G)$.
It includes the spherical swimmer ($G=0$) with $C_{0}^{\ast} = H^{\ast} = 2$.
For $C_0 < C_{0}^{\ast}$ the swimmer is swinging and for  $C_0 > C_{0}^{\ast}$ it is tumbling.

Linearizing the equations of motion [\refEqu{1}] around the fixed point $(\Psi=0,x=0)$
results in a harmonic oscillator equation
$\ddot{\Psi}+\vf(1-G)\Psi=0$ with
the eigen frequency $\omega_0=\sqrt{\vf(1-G)}$.
So, elongated swimmers swing with a smaller frequency around the centerline
compared to spherical swimmers.
The reason is that for elongated swimmers the reorientation rate or angular velocity is smaller compared
to spherical swimmers, when $|\Psi|<\pi/4$.
In particular, at fixed $x$ the angular velocity for
elongated swimmers  [\refEqu{1}, first line] depends on the orientation, 
$\dot{\Psi}=\Omega_y = \vf x ( 1-G\cos 2\Psi  )$,
in contrast to spherical swimmers.
We plot the angular velocity $\dot{\Psi}$ versus $\Psi$ in \refFig{rotation}(a).
For spherical swimmers at position $x$  it does not depend on 
orientation $\Psi$ (orange line)
while for ellipsoidal particles $\dot{\Psi}$ varies with $\Psi$ since
the strain rate
of the flow contributes to the angular velocity.
An elongated swimmer oriented parallel to the flow direction ($\Psi = 0$ or $\pi$)
rotates with a smaller $\dot{\Psi}$ as a spherical swimmer,
while at perpendicular orientation to the flow
 ($\Psi \approx \pm \pi/2$) the angular velocity is larger.
 In the extreme case of an infinitely thin needle ($G=1$)
 [blue curve in \refFig{rotation}(a)],
 the angular velocity vanishes at $\Psi = 0$ and  $\Psi = \pi$ and the needle
 just swims along the flow.
 We also sketch the reorientation rate
for spherical and elongated swimmers in \refFig{rotation}(b).

The fact that elongated swimmers rotate slower when oriented
in flow direction has a noticeable effect on the 
orientational distribution function of the swimmer,
\begin{equation}
 P(\Psi) \propto |\dot{\Psi}|^{-1} \propto (1-G\cos 2\Psi)^{-1}(C_0-1+g_0(\Psi,G))^{-1/2}
 \label{Eq:PPsi}
\end{equation}
To arrive at \refEq{PPsi}, we have eliminated location $x$ in the first line
of \refEqu{1} using the constant of motion $C_0$ of \refEq{H2D}.
In \refFigu{rotation2}(a) and (b), we show the respective probability distributions for spherical and ellipsoidal swimmers.
Spherical swimmers have the same probability distribution as the mathematical pendulum \cite{Baker06}.
$P(\Psi)$ assumes its absolute minimum where $\dot{\Psi}$ is largest.
This occurs at $\Psi=0$ when the swimmer has its maximum distance $|x|$ from the centerline
[see \refEqu{1}].
In the swinging motion  $P(\Psi)$ diverges at the centerline,
where $\dot{\Psi} = 0$ [green and red curve in \refFig{rotation2}(a)].
For the tumbling motion it stays finite since the swimmer does not reach the centerline
[blue curve in \refFig{rotation2}(a)].
For ellipsoidal swimmers $\dot{\Psi}$ also depends on the orientation $\Psi$.
Since they rotate slower at $\Psi = 0$ when they are oriented parallel to the flow,
an additional maximum at $\Psi = 0$ appears [\refFig{rotation2}(b)].

The reorientation rate $\dot{\Psi}$ also determines
the angular frequency of the swinging and tumbling motion,
\begin{equation}
\omega = \pi  \left( \int_{\Psi_{\text{min}}}^{\Psi_{\text{max}}} \frac{\mathrm{d}\Psi}{|\dot{\Psi}|} \right)^{-1}.
\label{Eq:333}
\end{equation}
For swinging motion, $\Psi_{\text{min}}$ and  $\Psi_{\text{max}}$ 
appear when $x=0$, so
\begin{equation}
\begin{split}
& \Psi_{\text{max}} =   -\Psi_{\text{min}} = \\
& \asec \left[ -\sqrt{8G/(G+1)} \coth \left( (C_0-1) \sqrt{2G(1 + G)}\right) \right],
\end{split}
\end{equation}
where $\asec$ is the arcus secans 
or the inverse function of $1/\cos x$.
For tumbling trajectories $ \Psi_{\text{min}} = 0$ and $ \Psi_{\text{max}} = \pi$.
We numerically integrate \refEq{333} for several geometry factors $G$
and plot it versus $C_0$ in \refFig{rotation2}(c).
For spherical swimmers the frequency decreases for larger oscillations or larger $C_0$
in the swinging state in full analogy to the mathematical pendulum [orange curve in \refFig{rotation2}(c)].
Larger oscillations mean that the swimmer orients more towards a perpendicular
orientation relative to the flow direction.
Since for elongated swimmers the reorientation rate is largest at $|\Psi| = \pi/2$,
the frequency $\omega$ first increases until it reaches a maximum for a specific swinging state
determined by a specific value of $C_0$
[blue curve in \refFig{rotation2}(c)].
In the tumbling state $\omega$ increases with $C_0$ for both the spherical and elongated swimmer.

\subsubsection{Three-dimensional motion}

 \begin{figure}[htb]
\includegraphics[width=.99\columnwidth]{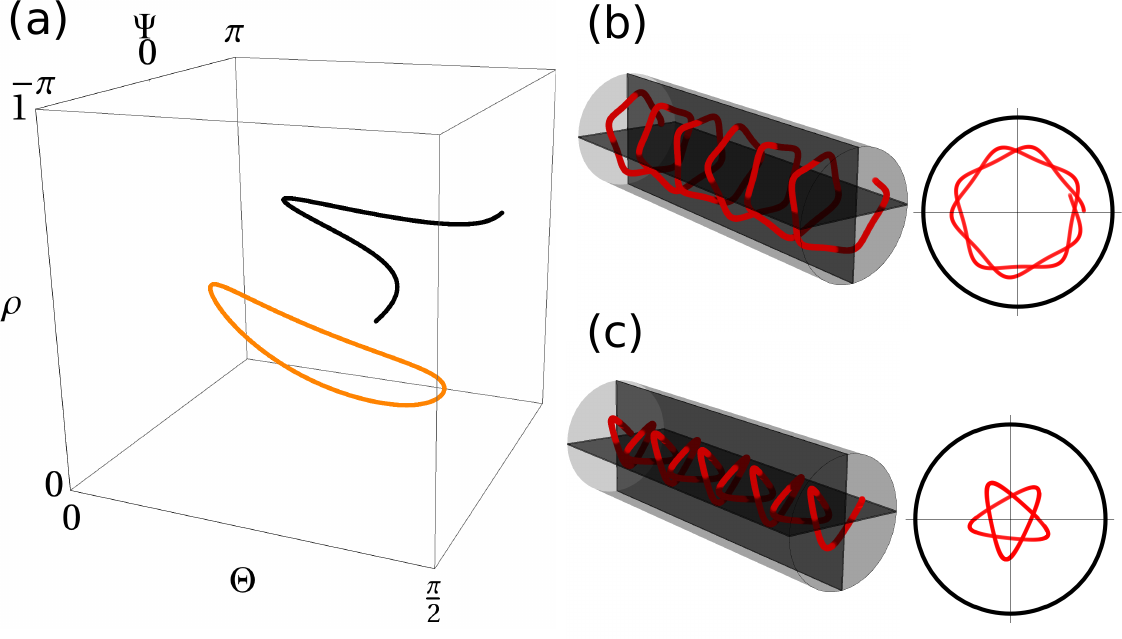}
\caption{
Swinging and tumbling motion for a microswimmer in a cylindrical tube for the flow strength $\bar{v}_{f} = 10$ and swimmer aspect ratio $\gamma=2$  $(G=0.6)$.
The initial conditions are $\Psi(0)=\pi / 4$, $\rho(0)=0.3$ and $\Theta(0)=\pi/5$
for swinging motion and $\Psi(0)=0$, $\rho(0)=0.7$ and $\Theta(0)=\pi/5$
for tumbling.
(a) Typical trajectories for swinging (orange)
and tumbling (black) motion in the $\rho$-$\Psi$-$\Theta$ phase space.
The trajectories $\mathbf{r}(t)$ in real space are shown in (b) for tumbling motion
and in (c) for swinging motion together with projections onto the channel cross section.
Note that the $z$ direction is squeezed relative to the lateral dimension.
}
\label{Fig:Cylindrical}
\end{figure}

For the three-dimensional motion [\refEqu{Plane3DG}] we are able to identify two constants of motion (see 
Appendix \ref{Sec:COM2D2}),
\begin{equation}
\begin{split}
\kappa_1 &= \tan^2\Theta / (1- G\cos 2 \Psi), \\
\kappa_2 &= \frac 1 2 \vf x^2 - g_1(\Psi,\Theta;G)
\label{Eq:Plane3DGCOM}
\end{split}
\end{equation}
with
\begin{equation} 
g_1(\Psi,\Theta; G) =  \frac{\atanh (\sqrt{2G/(1+G)}\cos\Psi\cos\Theta)}{\sqrt{2G(1+G)}}.
\end{equation}
The intersection of both constants $\kappa_1$ and $\kappa_2$ in the
$x$-$\Psi$-$\Theta$ phase space defines again periodic solutions of the ellipsoidal swimmer in the planar Poiseuille flow.
For $\Theta = 0$, $\kappa_1$ vanishes and $\kappa_2$ gives $C_0$.
We thus recover the two-dimensional motion discussed in the previous section.
In the full three-dimensional case, we can again divide the periodic solutions into swinging and tumbling motions.

Figure \ref{Fig:3DGraphs} illustrates specific examples for $\bar{v}_f = 10$
and $G=0.8$.
Figure \ref{Fig:3DGraphs}(a) shows the phases space curve for a typical swinging motion and
\refFig{3DGraphs}(d) for the swimmer in a tumbling state.
The initial conditions are $\Psi(0) = 0$ and $\Theta(0)= \pi/5$ for both cases,
and $x(0) = 0.3$ for (a) and  $x(0) = 0.7$ for (d).
So the constant of motion $\kappa_1$ is the same for both cases and it is illustrated by the orange
surfaces.
The second constant of motion $\kappa_2$ (green surfaces)
depends on $x$ and thereby
 defines if swinging or tumbling occurs.
The intersection between $\kappa_1$ and $\kappa_2$
is the actual trajectory in phase space (blue curve).
Figures \ref{Fig:3DGraphs}(b) and (e) show the motion of the
orientation vector $\vec{e}(t)$.
Since it is a unit vector, its arrowhead moves on the unit sphere $\mathbb{S}^2$.
For the swinging motion $\vec{e}$ moves back and forth with $e_y > 0$
on a trajectory that resembles a half ellipse (b), while it circles
around a closed trajectory during tumbling (d).
The closed trajectory is similar to the Jefferey orbit of tumbling passive
particles \cite{Jeffery22}, however for active particles $x$ is not constant.
In \refFig{3DGraphs}(c) and (f) the trajecories $\vec{r}(t)$ are shown.
For swinging motion the middle plane ($x=0$) is crossed periodically (c),
while tumbling motion always occurs on either side of the middle plane (f).
Both trajectories look two-dimensional tilted against the $z$ direction. Since the component 
$e_y(t)$ oscillates with time, they are clearly three-dimensional.

However, for spherical swimmers $G=0$ and $\dot{\Theta}=0$ [\refEq{Plane3DG}],
so that the component $e_y = \sin\Theta$ is constant in time.
The equation of motion for $\Psi$ is again the Pendulum equation but
it depends on the initial orientation $e_y = \const$,
\begin{equation}
\ddot{\Psi} + \xi \bar{v}_f \sin\Psi = 0,
\label{Eq:Plane3DEOMG0}
\end{equation}
where $\xi = \sqrt{1-e_y^2} = \text{const}$
is the length of the orientation vector in the $x$-$z$ plane.
A swimmer starting with $e_y \neq 0$
swings or tumbles at smaller frequency $\omega_0=\sqrt{\xi\bar{v}_f}$ compared to a swimmer with $e_y=0$
on a two-dimensional trajectory tilted against the $z$ direction.
So, the position $y(t)$ increases linearly in positive direction for $\Theta(0)>0$ or
in negative direction for $\Theta(0)<0$.
Since $\Theta=\text{const}$ the relevant phase space is the $x$-$\Psi$-space
similar to the two-dimensional problem.

 \begin{figure}[htb]
\includegraphics[width=.99\columnwidth]{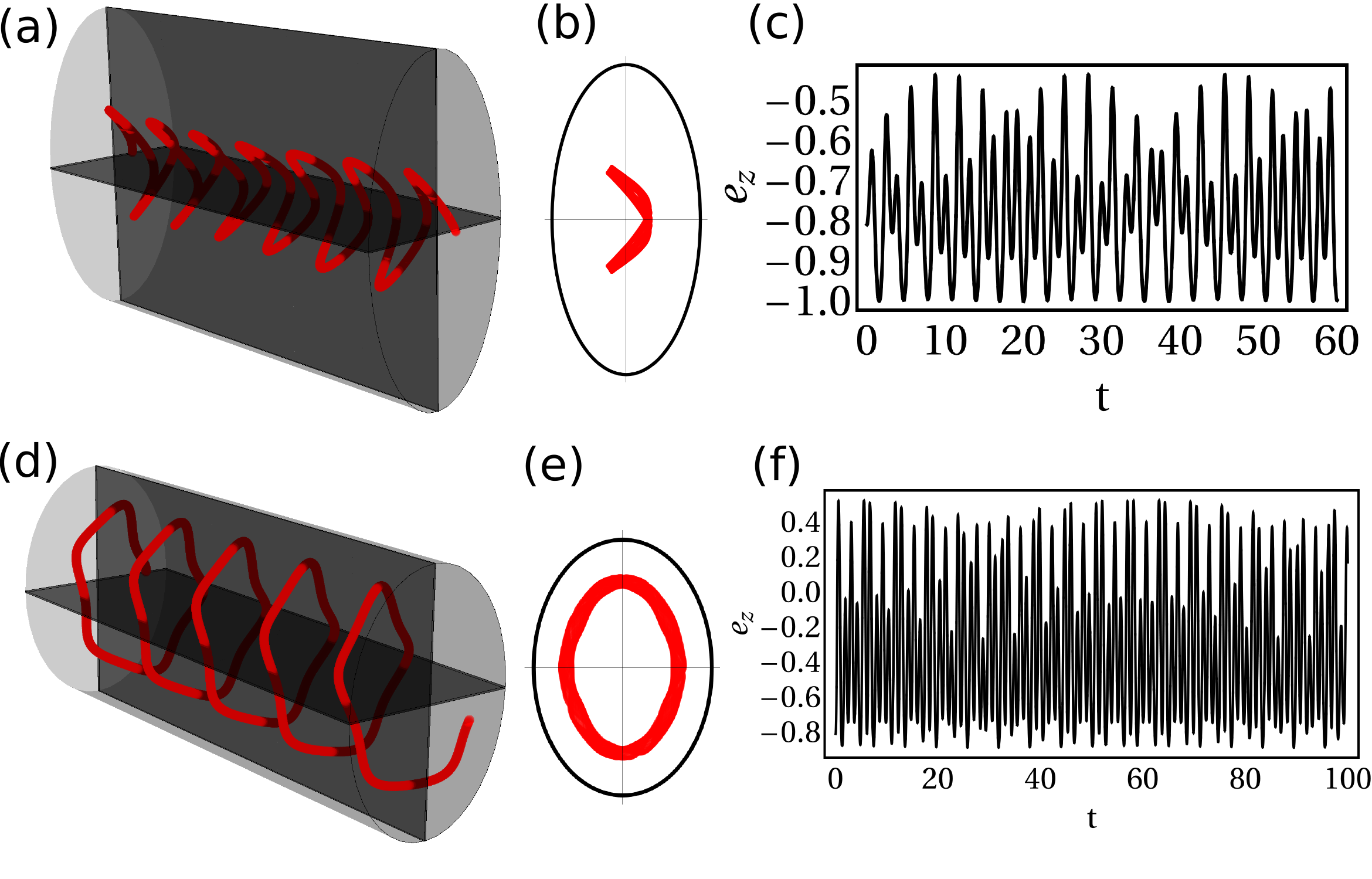}
\caption{(a)-(c) Two-frequency quasiperiodic swinging motion for the initial conditions
$\Psi(0)=0$, $\varphi(0)=0$, $\rho(0)=0.3$, $\Theta(0)=\pi/5$
 and parameters  $b=2$, $\gamma=2$, $vf=10$.
 (d)-(f) Tumbling motion for the  initial conditions
 $\Psi(0)=0$, $\varphi(0)=0$, $\rho(0)=0.7$, $\Theta(0)=\pi/5$
  and parameters $b=1.4$, $\gamma=2$ and $\vf=10$.
(a) and (d): Three-dimensional motion in the elliptical channel.
(b) and (e): Trajectories projected onto the cross section of the channel.
(c) and (f): Time evolution of the swimmer orientation component $e_z(t)$.
}
\label{Fig:QPSwing}
\end{figure}

\begin{figure*}[htb]
\resizebox{0.99\textwidth}{!}{\includegraphics{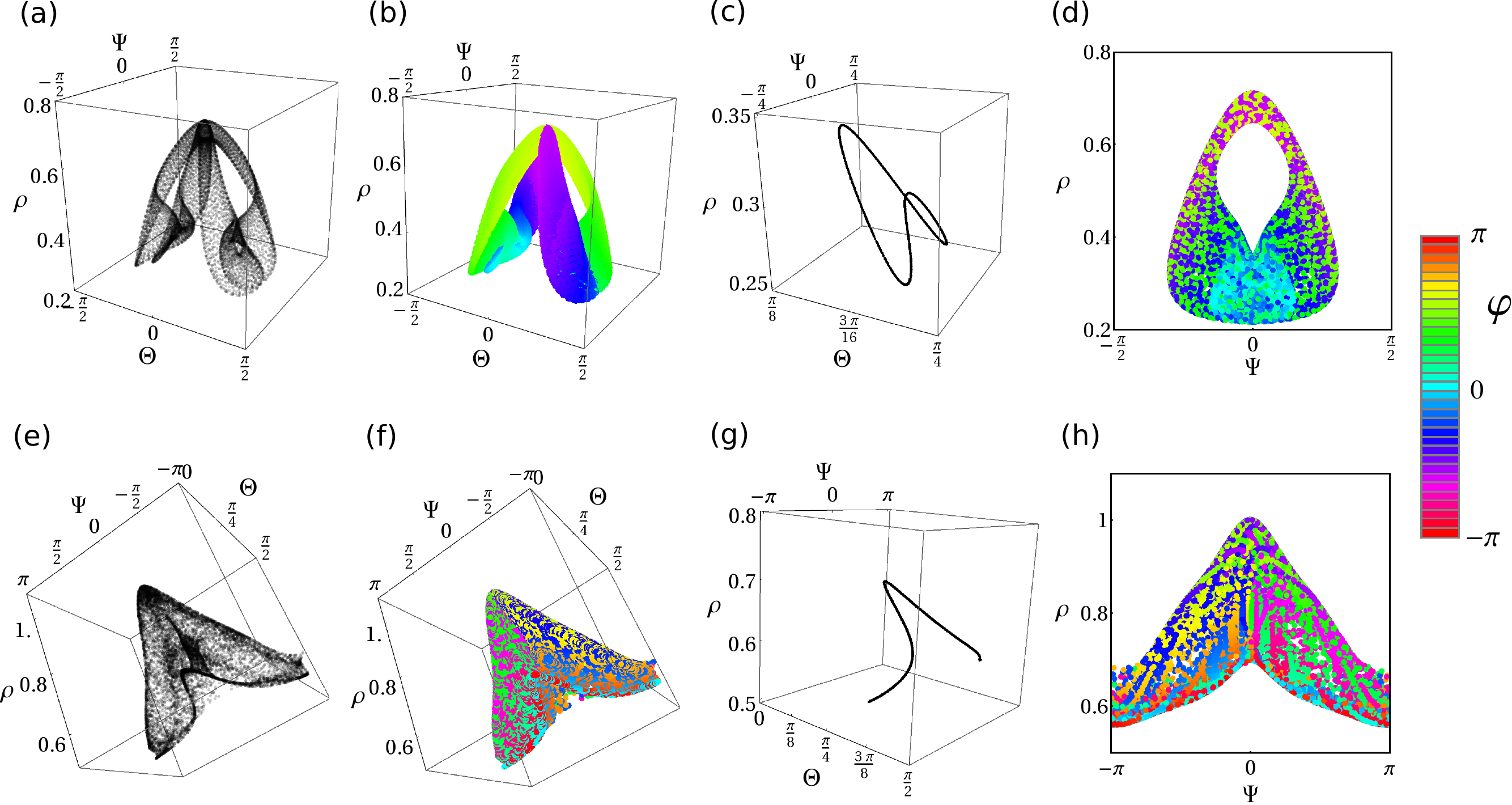}}
\caption{
Projected phase space and Poincar\'e sections for quasiperiodic swinging (a)-(d)
and tumbling (e)-(h) motion.
The initial conditions and system parameters are the same as in \refFig{QPSwing}.
The projections of the trajectories onto the $\rho$-$\Psi$-$\Theta$ phase space
lie on two-dimensional surfaces for swinging (a) and tumbling (e) motion.
(b) and (f): same trajectories as in (a) and (e) but with the coordinate $\varphi$ color-coded.
(c) and (g): Poincar\'e  sections at $\varphi=0$ ($\dot{\varphi}>0$).
One-frequency behavior in the Poincar\'e sections indicate two-frequency motion
in phase space \cite{Drazin94}.
(d) and (h):
Projections onto the $\rho$-$\Psi$ phase space
where $\varphi$ is color-coded.
}
\label{Fig:PSPM}
\end{figure*}

\subsection{Swimming in a cylindrical tube}

Now we discuss the motion of a swimmer in a cylindrical microchannel.
We use cylindrical coordinates $\rho,\varphi,z$ and express the 
orientation vector $\vec{e}$ in the cylindrical coordinate basis $(\hrho,\hphi,\hz)$,
$\eh(\Psi,\Theta) = e_{\rho}(\Psi,\Theta)\hrho + e_{\varphi}(\Theta)\hphi + e_z(\Psi,\Theta)\hz$
with
\begin{equation}
\erho = -\cos\Theta\sin\Psi, \ephi = \sin\Theta, \ez = -\cos\Theta\cos\Psi.
\label{Eq:erhophiz}
\end{equation}
Because of translational symmetry in $z$ direction and  rotational symmetry
with respect to $\varphi$, only the equations 
for $\rho$, $\Psi$ and $\Theta$ are coupled,
\begin{equation}
\begin{split}
\dot{\rho} &= -\cos\Theta\sin\Psi, \\
\dot{\Psi} &= \bar{v}_{f}\rho \left[1-G\cos 2\Psi  \right] - \sin\Theta\tan\Theta\cos\Psi/ \rho,  \\
\dot{\Theta} &= \left( 1 + 2 G \bar{v}_{f} \rho^2 \cos\Theta\cos\Psi  \right) \sin\Theta\sin\Psi / \rho.
\end{split}
\label{Eq:EOMClind}
\end{equation}
The relevant phase space  $\mathbb{R}\otimes\mathbb{S}^2$ is again three-dimensional.
For the spherical microswimmer ($G=0$) we already identified
two constants of motion. 
They lead to periodic trajectories with helical or helical-like 
swinging and tumbling motion \cite{Zoettl12}.
We derive the constants of motion
in Appendix~\ref{Sec:COMCyl}.
We were not able to construct constants of motion for an elongated swimmer in cylindrical Poiseuille flow.
Nevertheless, we numerically find that it performs again
periodic motions similar to the spherical swimmer.
Figure~\ref{Fig:Cylindrical}(a) shows two closed phase space orbits,
both for the swinging and tumbling state, for a swimmer
with aspect ratio $\gamma=2$ or geometry factor $G=0.6$.
We also illustrate the three-dimensional trajecories in the tube
and their two-dimensional projections onto the channel cross section 
Figs.~\ref{Fig:Cylindrical}(b) - (c).
For the swinging motion we have chosen initial conditions such that
the projection onto the cross-sectional plane is periodic.

\subsection{Quasi-periodic motion in a channel with elliptical cross section}

Now we consider the motion in a channel with elliptical cross section $(\bar{b}>1)$.
We again use a cylindrical coordinate system as for the cylindrical channel.
The symmetry in $\varphi$ direction no longer exists
and the relevant phase space is four-dimensional,
$ (\rho,\Psi,\Theta,\varphi) \in \mathbb{R}^2\otimes\mathbb{S}^2 $.
The equations of motion for the relevant coordinates read
\begin{equation}
\begin{split}
\dot{\rho} &= -\cos\Theta\sin\Psi, \\
\dot{\Psi} &= \bar{v}_{f}\rho \left(1-G\cos 2\Psi  \right)\left[  
\frac{\bar{b}^2+1}{2}-\frac{\bar{b}^2-1}{2}\cos 2\varphi
\right] \\
 &- \sin\Theta\tan\Theta\cos\Psi/ \rho  \\
 &- \bar{v}_{f}\rho \left( \frac{\bar{b}^2-1}{2}\right) (1+G)\sin 2\varphi \sin\Psi\tan\Theta,         \\
\dot{\Theta} &=
 \frac{\sin\Theta\sin\Psi}{\rho} \left[  1 +
  2 G \bar{v}_{f} \rho^2 \cos\Theta\cos\Psi(\cos^2\varphi + \bar{b}^2\sin^2\varphi) \right] \\
&- \bar{v}_{f}\rho \left( \frac{\bar{b}^2-1}{2} \right) \left( 1- G \cos 2 \Theta \right) \cos\Psi\sin 2 \varphi, \\
\dot{\varphi} &= \sin\Theta / \rho. \\
\end{split}
%\text{\dots insert equations of motion here \dots}.
\end{equation}
We again  solve them numerically and determine the swimmer trajectories.
Interestingly, both spherical and elongated swimmers
now follow quasiperiodic trajectories.
Figure~\ref{Fig:QPSwing} shows trajectories for a quasiperiodic swinging [Fig.\ \ref{Fig:QPSwing}(a)]
and tumbling motion [\refFig{QPSwing}(d)] 
and their projections onto the elliptical channel cross sections [\refFigu{QPSwing}(b) and (e)].
We also show the time evolution of the 
$z$ component of the orientation vector, $e_z(t)$,
 which shows two-frequency quasiperiodic behavior [\refFigu{QPSwing}(c) and (f)]
 as we will also demonstrate in \refFig{PSPM}. 
This indicates that the dynamics of a single trajectory takes place on a torus in the four-dimensional
phase space.
In \refFig{PSPM} we further illustrate the complex dynamics of the quasiperiodic swinging
[\refFigu{PSPM}(a) - (d)] and tumbling [\refFigu{PSPM}(e) - (h)] dynamics for the same
parameters as in \refFig{QPSwing}.
Since the quasiperiodic trajectories lie on a torus, they define a two-dimensional surface when projected onto the three-dimensional $\rho$-$\Psi$-$\Theta$ subspace of the full phase space.
Figures~\ref{Fig:PSPM}(a) and (e) illustrates these complex
surfaces.
In addition, we plot their Poincar\'e sections for $\varphi=0$, $\dot{\varphi}>0$
in \refFig{PSPM}(c) and (g).
The resulting closed curves confirm the two-frequency quasiperiodic motion on a torus \cite{Drazin94}.
In \refFig{PSPM}(b) and (f) we show the surfaces again with the azimuthal angle $\varphi$
color-coded.
For the swinging motion [\refFig{PSPM}(b)] one can clearly identify specific regions in the
$\rho$-$\Psi$-$\Theta$ space for each value of $\varphi$.
The projection of the surface onto the $\rho$-$\Psi$ plane 
[see \refFig{PSPM}(d)]
resembles the closed trajectories
of the swinging motion discussed earlier but now the curve is smeared out
in the $\rho$-$\Psi$ plane due to the quasiperiodicity.
The same is true for the tumbling motion, which we illustrate further in \refFigu{PSPM}(f) and (h).
The projected phase space trajectory in the $\rho$-$\Psi$ plane is a smeared-out version
of previous examples (\refFigu{PSPM}(h)).
However, the azimuthal angle $\varphi$ in the color-coded
surface of \refFigu{PSPM}(f) is distributed over the whole phase-space trajectory.

 Finally we note that we also studied the dynamics of spherical swimmers which also turned out to be  quasiperiodic.

\section{Conclusion} \label{sec.concl}

To conclude, we studied the two- and three-dimensional deterministic dynamics of
elongated swimmers in Poiseuille flow at low Reynolds number 
generalizing our work from Ref. \cite{Zoettl12}.
The state of the system is given by five variables, three for the position and two
for the orientation in the channel, respectively.
For all swimmers we find the same characteristic swimming states.
Depending on initial conditions and
parameters -- aspect ratio of the swimmer ($\gamma$),
geometry of the channel cross section ($\bar{b}$), and flow strength ($\bar{v}_f$) -- 
the microswimmer either swings with upstream orientation around the 
centerline of the channel
or it tumbles at sufficiently 
large
flow vorticities. 
The frequency of swinging and tumbling motion depends on the aspect ratio, 
in particular, it is smaller for elongated compared to spherical swimmers.
For the motion between two infinitely extended parallel plates and for the motion
in a cylindrical channel the relevant phase space is only three-dimensional
due to the symmetry of the channel cross section. 
In these cases we were able to identify two constants of motion which
results in a periodic motion visible in the three-dimensional phase space.
Interestingly, when we reduce the symmetry by choosing a channel with elliptical cross section, the phase space becomes four-dimensional and quasiperiodic motion occurs.

Note that in the framework of dynamical systems the motion of swimmers in Poiseuille
flow is not considered as dissipative. In laminar flow noninteracting swimmers move on separate
trajectories without aggregating, their dynamics does not reach a common fixpoint or limit cycle.
This changes when hydrodynamic interactions between a swimmer and the channel wall 
play a significant role as demonstrated in Ref. \cite{Zoettl12} where we explicitely
identified fixpoints and  limit cycles for the spherical swimmer. Other examples are
external forces such as gravitation  \cite{Durham09,Kessler85,Pedley92} or external light sources \cite{Peyla12} that influence the reorientation of the microswimmers.

\begin{appendix}
\section{Constants of motion}
\subsection{2D planar flow}
\label{Sec:COM2D}
A constant of motion can be found by deviding the first equation by the second equation
of \refEqu{1},
\begin{equation}
\frac{\dot{x}}{\dot{\Psi}} = \frac{d x}{d \Psi} = -\frac{\sin\Psi }{ \vf x (1-G\cos 2 \Psi)}.
\label{Eq:COM1}
\end{equation}
This expression can be separated and integrated,
\begin{equation}
\vf \int x \mathrm{d}x = -\int \frac{\sin\Psi \mathrm{d}\Psi}{1-G\cos 2\Psi},
\label{Eq:COM2}
\end{equation}
leading to
\begin{equation}
\frac{\vf}{2}x^2 =  \frac{\atanh\left( {\sqrt{2G/(1+G)}\cos\Psi  }\right) }{\sqrt{2G(1+G)}} + c
\label{Eq:COM3}
\end{equation}
where $c$ is a constant and $C_0=c+1$ is the constant of motion 
defined in \refEq{H2D}.

\subsection{3D planar flow}
\label{Sec:COM2D2}
The first constant of motion $\kappa_1$ is identified by calculating
\begin{equation}
\frac{\dot{\Theta}}{\dot{\Psi}} = \frac{d \Theta}{d \Psi} =
 -\frac{G}{2}\frac{\sin 2\Theta\sin 2\Psi }{ 1-G\cos 2 \Psi}.
\label{Eq:COM4}
\end{equation}
We again separate the variables and integrate,
\begin{equation}
\int \frac{\mathrm{d}\Theta}{G\sin 2\Theta}  =
 \int \frac{\sin 2\Psi \mathrm{d}\Psi}{1-G\cos 2\Psi},
\label{Eq:COM5}
\end{equation}
resulting in
\begin{equation}
\frac{1}{2G}[\log(1-G\cos 2\Psi)] + c = \frac{1}{G}[\log(\tan\Theta)].
\label{Eq:COM6}
\end{equation}
Taking the exponential of this expression and
defining $\kappa_1$ as $\kappa_1 = e^{-2c}$
results in the first constant of motion in \refEqu{Plane3DGCOM}.
Using $\kappa_1$, 
eliminating $\Theta$  %by using $\kappa_1$
and calculating
\begin{equation}
\frac{\dot{\Psi}}{\dot{x}} = \frac{d \Psi}{d x} =
 -\frac{\bar{v}_f x}{\sin\Psi} \left( 1-G\cos 2 \Psi \right) \sqrt{1+\kappa^{-1}(1-G\cos 2 \Psi)},
\label{Eq:COM7}
\end{equation}
after separation of variables and integration then leads to
\begin{equation}
\frac{\bar{v}_f}{2} x^2 = \frac{\atanh (\sqrt{2G/(1+G)}\cos\Psi\cos\Theta)}{\sqrt{2G(1+G)}}
+c
\label{Eq:COM6}
\end{equation}
where we defined $\kappa_2$ as $\kappa_2=1+c$ as
the second constant of motion in \refEqu{Plane3DGCOM}.

\subsection{Cylindrical flow for $G=0$}
\label{Sec:COMCyl}
The equations of motion for spherical particles swimming in a cylindrical Poiseuille flow
are given by \refEqu{EOMClind} but with $G=0$.
Using \refEqu{EOMClind} the first constant of motion for a spherical swimmer is determined by taking
\begin{equation}
\frac{\dot{\Theta}}{\dot{\rho}} = \frac{d \Theta}{d \rho} = -\tan\Theta / \rho,
\label{Eq:COM8}
\end{equation}
after separation of variables and integration leading to
\begin{equation}
-\log(\sin\Theta) = \log(\rho) + c,
\label{Eq:COM9}
\end{equation}
and to the first constant of motion $L = e^{-c} = \rho\sin\Theta$ 
 which is proportional to the angular momentum of the swimmer in the $z$ direction.
The second constant of motion  is identified by taking
\begin{equation}
\frac{\dot{\Psi}}{\dot{\Theta}} = \frac{d \Psi}{d \Theta} =
\frac{\bar{v}_{f} L^2}{\sin^3\Theta\sin\Psi} - \frac{\tan\Theta}{\tan\Psi}.
\label{Eq:COM10}
\end{equation}
This nonlinear differential equation can be transformed into 
a linear equation by using the ansatz $w=\cos\Psi$ leading to
\begin{equation}
\frac{d w}{d \Theta} + w\tan\Theta = - \frac{\bar{v}_{f} L^2}{\sin^3\Theta}.
\label{Eq:COM11}
\end{equation}
The solution is 
\begin{equation}
w=\cos\Psi = \frac{L^2 \bar{v}_f}{2\sin\Theta\tan\Theta} - \frac{c}{\cos\Theta}
\label{Eq:COM12}
\end{equation}
and we define the second constant of motion as
\begin{equation}
M = c + \frac{1}{2} L^2 \bar{v}_{f} + 1 = \frac{1}{2} \bar{v}_{f} \rho^2 + 1 - \cos\Theta\cos\Psi.
\label{Eq:COM12}
\end{equation}

\end{appendix}

\end{document}